\documentclass[sn-nature]{sn-jnl}

\usepackage{graphicx}%
\usepackage{multirow}%
\usepackage{amsmath,amssymb,amsfonts}%
\usepackage{amsthm}%
\usepackage[title]{appendix}%
\usepackage{xcolor}%
\usepackage{textcomp}%
\usepackage{manyfoot}%
\usepackage{booktabs}%
\usepackage{algorithm}%
\usepackage{algorithmicx}%
\usepackage{algpseudocode}%
\usepackage{listings}%
\usepackage{etoolbox}

\theoremstyle{thmstyleone}%
\theoremstyle{thmstyletwo}%

\theoremstyle{thmstylethree}%

\newcommand\actaa{Acta Astronom.}
\newcommand\jcap{J. Cosmol. Astropart. P.}
\newcommand\prl{Phys. Rev. Lett.}

\newcommand\prx{Phys. Rev. X}
\newcommand\pdu{Phys. Dark Universe}
\newcommand\apj{Astrophys. J.}
\newcommand\aj{Astron. J.}
\newcommand\pasp{Publ. Astron. Soc. Pac.}
\newcommand\mnras{Mon. Not. R. Astron. Soc.}
\newcommand\aap{Astron. Astrophys.}
\newcommand\apjl{Astrophys. J. Lett.}
\newcommand\apjs{Astrophys. J. Suppl. S.}
\newcommand\nat{Nature}

\newcommand\tE{t_{\rm E}}
\newcommand\pirel{\pi_{\rm rel}}
\newcommand\piE{\pi_{\rm E}}

\begin{document}

\title[Article Title]{No massive black holes in the Milky Way halo}

\author*[1]{\fnm{Przemek} \sur{Mr\'oz}}\email{pmroz@astrouw.edu.pl}
\author[1]{\fnm{Andrzej} \sur{Udalski}}
\author[1]{\fnm{Micha\l{} K.} \sur{Szyma\'nski}}
\author[1]{\fnm{Igor} \sur{Soszy\'nski}}
\author[1]{\fnm{\L{}ukasz} \sur{Wyrzykowski}}
\author[1]{\fnm{Pawe\l{}} \sur{Pietrukowicz}}
\author[1]{\fnm{Szymon} \sur{Koz\l{}owski}}
\author[1]{\fnm{Rados\l{}aw} \sur{Poleski}}
\author[1]{\fnm{Jan} \sur{Skowron}}
\author[1]{\fnm{Dorota} \sur{Skowron}}
\author[1,2]{\fnm{Krzysztof} \sur{Ulaczyk}}
\author[1]{\fnm{Mariusz} \sur{Gromadzki}}
\author[1,3]{\fnm{Krzysztof} \sur{Rybicki}}
\author[1]{\fnm{Patryk} \sur{Iwanek}}
\author[1]{\fnm{Marcin} \sur{Wrona}}
\author[1]{\fnm{Milena} \sur{Ratajczak}}

\affil*[1]{\orgdiv{Astronomical Observatory}, \orgname{University of Warsaw}, \orgaddress{\street{Al. Ujazdowskie 4}, \city{Warszawa}, \postcode{00-478}, \country{Poland}}}
\affil[2]{\orgdiv{Department of Physics}, \orgname{University of Warwick}, \orgaddress{\street{Coventry CV4 7 AL}, \country{UK}}}
\affil[3]{\orgdiv{Department of Particle Physics and Astrophysics}, \orgname{Weizmann Institute of Science}, \orgaddress{\street{Rehovot 76100}, \country{Israel}}}



\maketitle

\textbf{The gravitational wave detectors have unveiled a population of massive black holes that do not resemble those observed in the Milky Way \cite{ligo_virgo_2016,ligo_virgo_2019,ligo_virgo_2021} and whose origin is debated \cite{belczynski2016,askar2017,rodriguez2018}. According to one possible explanation, these black holes may have formed from density fluctuations in the early Universe (primordial black holes) \cite{bird2016,sasaki2016,clesse2017}, and they should comprise from several to 100\% of dark matter to explain the observed black hole merger rates \cite{carr2021b,jedamzik2021,escriva2023}. 
If such black holes existed in the Milky Way dark matter halo, they would cause long-timescale gravitational microlensing events lasting years \cite{paczynski1986}. The previous experiments were not sufficiently sensitive to such events \cite{alcock2000b,tisserand2007,wyrzyk1,moniez2022}.
Here we present the results of the search for long-timescale microlensing events among the light curves of nearly 80 million stars located in the Large Magellanic Cloud that were monitored for 20 years by the OGLE survey \cite{udalski2015}. 
We did not find any events with timescales longer than one year, whereas all shorter events detected may be explained by known stellar populations. We find that compact objects in the mass range from $1.8 \times 10^{-4}$ to $6.3\,M_{\odot}$ cannot compose more than 1\% of dark matter, and those in the mass range from $1.3 \times 10^{-5}$ to $860\,M_{\odot}$ cannot make up more than 10\% of dark matter. Thus, primordial black holes in this mass range cannot simultaneously explain a significant fraction of dark matter and gravitational wave events.}

We analyze the photometric observations of the Large Magellanic Cloud (LMC) that have been collected for nearly 20 years by the Optical Gravitational Lensing Experiment (OGLE), during its third (OGLE-III; 2001--2009; \cite{udalski2003}) and fourth (OGLE-IV; 2010--2020; \cite{udalski2015}) phases. Since OGLE-III and OGLE-IV had similar observing setups, it was possible to merge the observations to create a 20-year-long photometric time-series data set. We developed a new method of reductions of photometric observations, which enabled us to obtain homogeneous light curves. The design of the survey, extraction of photometry, and methods used to search for microlensing events and calculate the event detection efficiency are described in detail in a companion paper \cite{mroz2024a}.

About 33 million objects are detected in the overlapping OGLE-III/OGLE-IV region, and an additional 29 million objects are observed by OGLE-IV only. The number of stars that may be microlensed is even higher because of blending, which occurs when two or more stars cannot be resolved in ground-based seeing-limited images. We used the archival high-resolution images from the Hubble Space Telescope \cite{holtzman2006} to correct the star counts for blending. After removing the contribution from foreground Milky Way stars, we found that the survey monitored for microlensing about 78.7\,million source stars in the LMC brighter than $I=22$\,mag \cite{mroz2024a}.

We searched for microlensing events using a variation of the method described by \cite{mroz2017}. The algorithm tries to identify a flat portion of the light curve, and then searches for consecutive data points that are magnified with respect to the flat part. Then, a standard point-source point-lens microlensing model \cite{paczynski1986} is fitted to the light curve, and the goodness-of-the-fit statistics are evaluated. The events are selected on the basis of a series of selection cuts. This procedure enabled us to find thirteen events that fulfill all detection criteria. Additionally, three events were identified by a manual inspection of the light curves, although they did not meet all selection criteria. The sample of thirteen events is used for a later statistical analysis \cite{mroz2024a}.

We also carried out extensive light curve simulations to measure the event detection efficiency as a function of the event timescale \cite{mroz2024a}. To this end, we created synthetic light curves of microlensing events by injecting the microlensing signal into the light curves of constant stars observed by the project. Then, we measured the fraction of simulated events that passed all selection criteria. This procedure enabled us to take into account the noise in the data, as well as the effects of irregular sampling, gaps in the data, outliers, etc.

We parameterize the fraction of the dark matter in the form of primordial black holes (PBHs) and other compact objects as $f=M_{\rm PBH}/M_{\rm DM}$, where $M_{\rm DM}$ is the total mass of the dark matter halo, and $M_{\rm PBH}$ is the total mass of PBHs in it \cite{wyrzyk1}. We note that we expect to detect some gravitational microlensing events even if $f=0$: they come from lensing objects located in the Milky Way disk and the LMC itself. The latter phenomenon is also called self-lensing \cite[e.g.,][]{sahu1994}.

We found that all thirteen microlensing events detected in our survey can be explained by brown dwarfs, stars, and stellar remnants located in the LMC and the foreground Milky Way disk (Methods). We adopted the LMC model from ref. \cite{alcock2000b} with some modifications, and considered two Milky Way disk models by \cite{cautun2020} and \cite{han_gould2003} (Methods). We calculated the theoretical microlensing event rate and the event timescale distribution in each field analyzed by OGLE. Then, we estimated the expected number of events by multiplying the event rate by the duration of the observations, the number of source stars, and the average event detection efficiency. In total, we expected to find 5.7 events due to LMC lenses and 7.0 or 14.7 events due to lenses in the Milky Way disk (depending on the adopted model), which can be compared to the total number of thirteen events in the final statistical sample. We also found that positions, timescales, and microlensing parallaxes of the detected events are consistent with the predictions of the adopted model (Methods).

To infer the constraints on the value of $f$, we calculated the expected number of events and their timescale distribution, assuming that the entire dark matter halo is composed of compact objects of the same mass $M$, and taking into account the measured event detection efficiency \cite{mroz2024a}. We assumed the contracted Milky Way halo model by \cite{cautun2020} and the LMC halo model by \cite{erkal2019} (Methods). The expected distributions of event timescales, for three example PBH masses (0.01, 1, and $100\,M_{\odot}$) are presented in Extended Data Figure~\ref{fig:n_exp}a. The mean event timescales approximately scale as $\sqrt{M}$ and amount to 8, 70, and 600 days, respectively. On the other hand, the theoretical event rate is inversely proportional to $\sqrt{M}$. 

Our experiment has the highest sensitivity to PBHs with masses of $0.01\,M_{\odot}$; we should have detected more than 1100 events if the entire dark matter were composed of such objects. However, thanks to the long duration of observations, OGLE has a high sensitivity for even more massive objects. For $M=1\,M_{\odot}$, we should have detected 554 events; for $M=10\,M_{\odot}$---258 events; for $M=100\,M_{\odot}$---99 events; for $M=1000\,M_{\odot}$---27 events. Additional information that can be used to further constrain the abundance of PBHs in the Milky Way and LMC haloes is included in the timescales of observed events (Methods).

\begin{figure*}[bh]
\centering
\includegraphics[width=\textwidth]{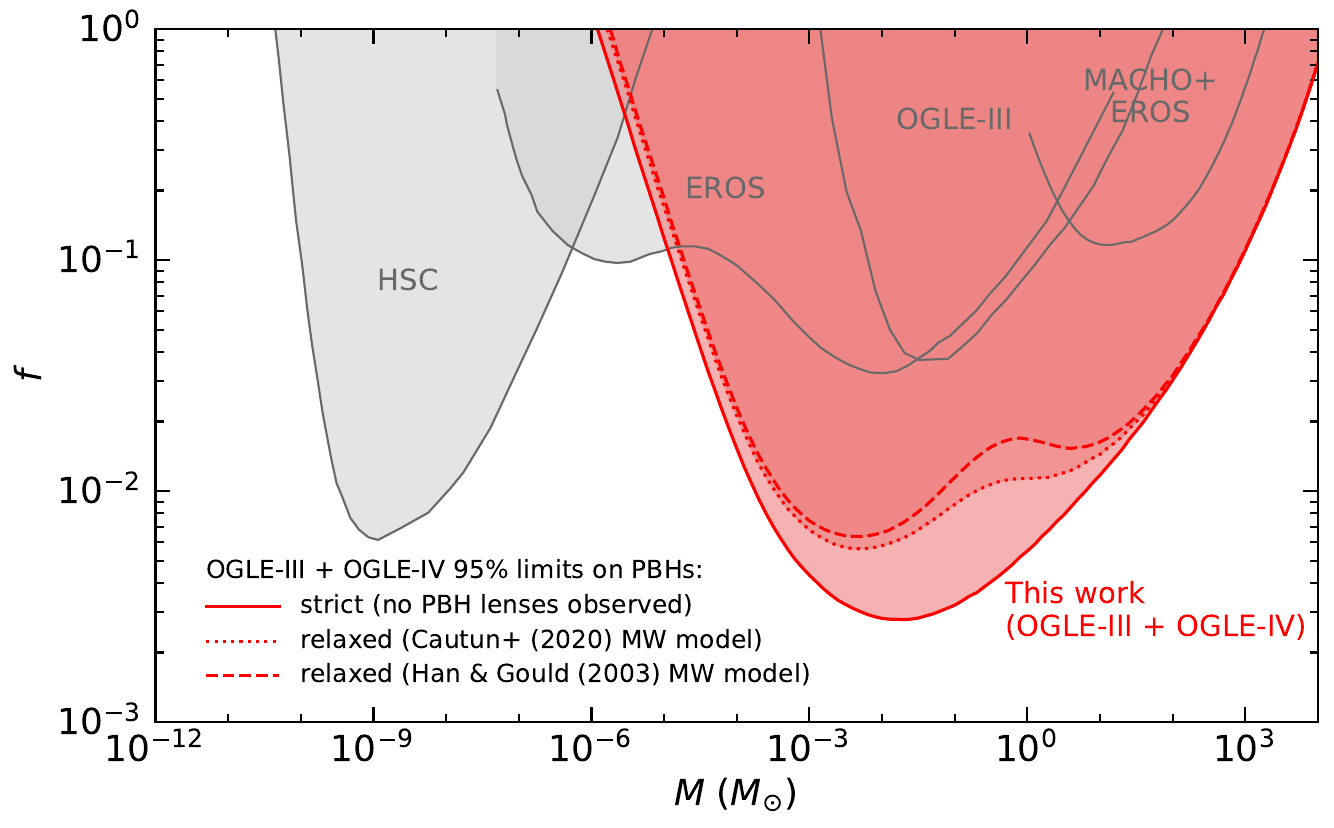}
\caption{\textbf{95\% upper limits on PBHs (and other compact objects) as constituents of dark matter.} The solid red line marks the limits derived in this paper under the assumption that all gravitational microlensing events detected by OGLE in the direction of the LMC are due to objects in the LMC itself or the Milky Way disk. If this assumption is relaxed, the limits (dotted and dashed lines) depend on the choice of the LMC and Milky Way disk model (\cite{cautun2020} or \cite{han_gould2003}, respectively). The gray lines mark the limits determined by the following surveys: EROS \cite{tisserand2007}, OGLE-III \cite{wyrzyk1}, Hyper Suprime-Cam (HSC) \cite{niikura2019}, and MACHO+EROS \cite{moniez2022}. The new limits are available online as Supplementary Table~1.}
\label{fig:bounds}
\end{figure*}

Our 95\% upper limits on PBHs (and other compact objects) as constituents of dark matter are presented in Figure~\ref{fig:bounds}. The solid red line marks the strict limits derived in this paper under the assumption that all gravitational microlensing events detected by OGLE in the direction of the LMC are due to known stellar populations in the LMC itself or the Milky Way disk. These limits are inversely proportional to the number of events expected if the entire dark matter was composed of compact objects of a given mass. As expected, the limits are strongest ($f = 2.8\times10^{-3}$) for $M \approx 0.01\,M_{\odot}$, for which the model predicts the largest number of expected events. The PBHs of mass $M=1\,M_{\odot}$ may make up less than $f=0.55\%$ of dark matter; $M=10\,M_{\odot}$---$f=1.2\%$; $M=100\,M_{\odot}$---$f=3.0\%$; $M=1000\,M_{\odot}$---$f=11\%$.
The PBHs in the mass range $1.8 \times 10^{-4}\,M_{\odot} < M < 6.3\,M_{\odot}$ cannot compose more than 1\% of dark matter, and the PBHs in the mass range $1.3 \times 10^{-5}\,M_{\odot} < M < 8.6 \times 10^2\,M_{\odot}$ cannot make up more than 10\% of dark matter.

We also found that, thanks to the long duration of the survey and the large number of stars monitored, the derived limits weakly depend on the choice of the Milky Way halo model. In particular, we tested the dark matter halo model of \cite{jiao2023} and found consistent results (as shown in Extended Data Figure~\ref{fig:comparison}a). The latter model was based on recent measurements of the rotation curve in the outer regions of the Milky Way by the \textit{Gaia} satellite.

The dotted and dashed lines in Figure~\ref{fig:bounds} mark the relaxed limits, for the derivation of which we did not make any assumptions about the origin of events. These limits only slightly depend on the choice of the Milky Way disk model. 
Overall, the differences between different models are minimal for the least massive ($M \lesssim 10^{-4}\,M_{\odot}$) and the most massive ($M \gtrsim 10\,M_{\odot}$) PBHs, because we do not expect that known stellar objects would produce microlensing events with timescales that could be attributed to those by extremely low- or high-mass PBHs. In the intermediate mass range ($10^{-4}\,M_{\odot} \lesssim M \lesssim 10\,M_{\odot}$), microlensing events caused by PBHs may be mistaken with those caused by known stellar populations in the LMC or the Milky Way disk. Therefore, the relaxed limits are slightly weaker than the strict ones in this mass range.

Limits presented in Figure~\ref{fig:bounds} are derived for a delta-function mass function of PBHs and, in principle, may become weaker if the underlying mass function is extended. Such extended mass functions of PBHs are frequently discussed in the context of binary black hole mergers discovered by gravitational wave detectors \cite{jedamzik2021,carr2021b,escriva2023}. For example, the model presented by \cite{carr2021b} predicts four peaks in the mass spectrum of PBHs at $10^{-6}$, $1$, $30$, and $10^{6}\,M_{\odot}$ (Extended Data Figure~\ref{fig:carr_model}a). These peaks would be associated with different phase transitions in the quark--gluon plasma filling the early universe ($W^{\pm}/Z^0$ decoupling, the quark--hadron transitions, and $e^+e^-$ annihilation), which are thought to enhance the formation of PBHs.

Several studies argue that such a multi-peak mass function of PBHs can naturally explain the observed merger rates of black hole binaries observed by gravitational wave detectors and a significant fraction (from several to 100\%) of dark matter \cite[e.g.,][]{jedamzik2021,carr2021b,escriva2023}. This hypothesis has one important prediction: PBHs in the Milky Way dark matter halo should cause long-timescale gravitational microlensing events that may last years. For example, if the entire dark matter was composed of PBHs with the mass spectrum described by the model of \cite{carr2021b}, we should have detected over 500 microlensing events (Extended Data Figure~\ref{fig:carr_model}b). The non-detection gives us a 95\% upper limit on the fraction of dark matter in the form of PBHs of $f=1.2\%$ (assuming a Milky Way disk model by \cite{cautun2020}). Similar limits of the order of 1\% can be obtained for other multi-peak mass functions proposed in the literature. 
Our observations, therefore, demonstrate that PBHs with masses in the range $1 \lesssim M \lesssim 1000\,M_{\odot}$ cannot comprise a significant fraction of the dark matter and, at the same time, explain the observed black hole merger rates as was argued by refs. \cite[][]{jedamzik2021,carr2021b,escriva2023}.

Our observations provide limits on any compact objects that may compose dark matter, not just PBHs. In particular, neutron stars and stellar-origin black holes are estimated to make up less than $0.01\%$ of the total Milky Way halo mass each \cite{olejak2020,abramowicz2022}, well below the limits presented in this paper.

\newpage

\section*{Methods}

\subsection*{Model}
\label{sec:model}

Let us suppose that we have a sample of $N_{\rm obs}$ events and let $t_{{\rm E},i}$ be the Einstein timescale of the $i$th event. We use that information to measure $f$ (or derive upper limits on $f$) by maximizing the likelihood function defined as
\begin{equation}
\mathcal{L}(f, M)=e^{-N_{\rm exp}(f,M)}\prod_{i=1}^{N_{\rm obs}}\left[N_{\rm s} \Delta t  \frac{d\Gamma}{dt_{\rm E}}\left(t_{{\rm E},i},f,M\right)\varepsilon(t_{{\rm E},i})\right],
\label{eq:likelihood}
\end{equation}
where $N_{\rm s}$ is the number of microlensing sources observed in the experiment, $\Delta t$ is the duration of observations, and $\varepsilon(\tE)$ is the event detection efficiency in the experiment (as a function of the Einstein timescale). Here $d\Gamma/d\tE$ is the differential event rate, which contains contributions from lenses in the Milky Way disk, LMC, and Milky Way and LMC dark matter haloes:
\begin{align}
\begin{split}
\frac{d\Gamma}{d\tE} &= f \frac{d\Gamma_{\rm MW\ halo}}{d\tE} \left(\tE, M\right) + f \frac{d\Gamma_{\rm LMC\ halo}}{d\tE} \left(\tE, M\right)\\
&+ \frac{d\Gamma_{\rm MW\ disk}}{d\tE} \left(\tE\right) + \frac{d\Gamma_{\rm LMC}}{d\tE} \left(\tE\right),
\end{split}
\end{align}
and
\begin{equation}
N_{\rm exp}(f, M) = N_{\rm s} \Delta t \int \frac{d\Gamma}{d\tE}\left(t_{{\rm E}}',f,M\right) \varepsilon(\tE') d\tE'
\label{eq:n_exp}
\end{equation}
is the expected number of events. Eq.~(\ref{eq:likelihood}) can be derived by dividing the observed event timescale distribution into infinitesimally small bins that contain either one or zero events, and assuming that the number of events detected in each bin follows the Poisson distribution. The microlensing event rate and the event timescale distribution are calculated following the standard approach \citep[e.g.,][]{clanton2014}. We describe the components of the model in the following subsections. 

\subsubsection*{Milky Way halo}
\label{sec:mw_halo}

Our model is based on the contracted halo model of \cite{cautun2020}, which was inferred by fitting physically motivated models to the \textit{Gaia} DR2 Galactic rotation curve and other data \cite{eilers2019}. The model includes the effect of the contraction of the dark matter halo in the presence of baryons, which was observed in large galaxy formation simulations \citep[e.g.,][]{schaye2015,fattahi2016,grand2017}. The model predicts the total mass of the dark matter Milky Way halo of $0.97 \times 10^{12}\,M_{\odot}$ within 200 kpc.

We use the best-fitting model of \cite{cautun2020} to predict the rotation curve of the Milky Way $V_0(R)$ (which, by design, matches the \textit{Gaia} data very well), where $R$ is the Galactocentric radius. We assume that the distribution of velocities of halo particles (in the rest frame of the Galaxy) can be considered as Maxwellian \cite{griest1991} with the standard deviation of velocities in one direction equal to $V_0(R)/\sqrt{2}$. (There is some evidence that (at least some) black holes may attain a large recoil velocity after a binary black hole merger \cite[e.g.,][]{varma2022}. If all PBHs had such additional velocity, the derived limits would be modified, such that they would be moved toward more massive PBHs.) Assuming that the Milky Way halo is composed of objects of identical mass $M$, the mean Einstein timescale of events predicted by the model is equal to $\tE = 62\,\mathrm{d}\sqrt{M/M_{\odot}}$, which is in good agreement with the predictions of models by \cite{alcock2000b} or \cite{moniez2022}.

\subsubsection*{LMC halo}

The \textit{Gaia} observations indicate that stars in the Orphan stream have velocity vectors that are significantly misaligned with the stream track \cite{koposov2019}. Ref.~\cite{erkal2019} demonstrated that this effect can be explained by gravitational perturbations from the LMC, and inferred the total mass of the LMC of $1.49\times 10^{11}\,M_{\odot}$ (in the model with the spherical Milky Way, which can move in response to the LMC; ref.~\cite{erkal2019} also considered models with an oblate or a prolate Milky Way halo, which result in the LMC mass that is $5-8\%$ smaller). The total mass of the Milky Way in the \cite{erkal2019} model (enclosed within a radius of 50\,kpc) is $4.04 \times 10^{11}\,M_{\odot}$, which is in excellent agreement with that inferred from the \cite{cautun2020} model of the Milky Way halo ($4.1 \times 10^{11}\,M_{\odot}$).

Following \cite{erkal2019}, we assume that the LMC halo can be modeled by a Hernquist profile with the total mass of $1.49\times 10^{11}\,M_{\odot}$. The scale length is 17.1\,kpc, and is taken so that the mass enclosed within 8.7\,kpc matches the measured value of $1.7\times10^{10}\,M_{\odot}$ from \cite{van_der_marel2014}.

\subsubsection*{Milky Way disk}
\label{sec:mw_disk}

For consistency with the Milky Way halo model presented above, we adopt the best-fitting thin and thick disk models from \cite{cautun2020}. The stellar density is described by the double exponential profile with a scale height of 0.3\,kpc (thin disk) or 0.9\,kpc (thick disk) and a radial scale length of 2.63\,kpc (thin disk) or 3.80\,kpc (thick disk). The total masses of the thin and thick disks are $3.18 \times 10^{10}\,M_{\odot}$ and $0.92\times 10^{10}\,M_{\odot}$, respectively. As an alternative, we also consider the thin and thick Milky Way disk models of \cite{han_gould2003}. The scale heights in the model are 0.156\,kpc (thin disk) and 0.439\,kpc (thick disk), the scale length is 2.75\,kpc for both think and thick disk. The total masses of the thin and thick disks are $0.98 \times 10^{10}\,M_{\odot}$ and $0.60\times 10^{10}\,M_{\odot}$, respectively. We assume that stars in the Milky Way disk follow the circular velocity curve derived above with a dispersion of 30\,km\,s$^{-1}$ in each direction. We adopt the mass function of \cite{kroupa2001}.

\subsubsection*{LMC}
\label{sec:lmc_disk}

To take into account the effects of self-lensing by stars located within the LMC itself, we adopt the LMC disk and bar models of \cite{gyuk2000} and \cite{alcock2000b} with some slight modifications. The disk is modeled as a double exponential profile with the scale length of 1.8\,kpc and the scale height of 0.3\,kpc \cite{gyuk2000}, the inclination and position angle of $25^{\circ}$ and $132^{\circ}$, respectively \cite{pietrzynski2019}. The bar is modeled as a triaxial Gaussian with the scale lengths of $x_b = 1.2$, $y_b=z_b = 0.44$\,kpc along the three axes \cite{gyuk2000,jetzer2002,sajadian2021}. We adopt the total stellar mass (disk + bar) of the LMC of $2.7\times 10^9\,M_{\odot}$ \cite{van_der_marel2002}, and the bar-to-disk mass ratio of 0.2 \cite{mancini2004,calchi_novati2009}. We adopt the LMC distance of 49.59\,kpc \cite{pietrzynski2019}. We assume that the stellar mass function is identical to that of the Milky Way. We use an empirical model of the kinematics of stars in the LMC, which is based on \textit{Gaia}~EDR3 data (see below). 

\subsection*{Limits on the PBH abundance}
\label{sec:limits}

\subsubsection*{Limits under the hypothesis that all microlensing events detected by OGLE can be explained by known stellar populations}
\label{sec:limits1}

We start by deriving upper limits on the PBH abundance by assuming that all detected events can be explained by the known stellar populations located in the Milky Way disk or the LMC itself. In other words, we assume that we did not see any events due to dark matter in the form of compact objects. In this case, the likelihood function (Eq.~(\ref{eq:likelihood})) simplifies to:
\begin{equation}
\mathcal{L}(f, M)=e^{-N_{\rm exp}(f,M)},
\label{eq:likelihood2}
\end{equation}
where $N_{\rm exp}(f,M)$ is the expected number of microlensing events, defined in Eq.~(\ref{eq:n_exp}).

Extended Data Figure~\ref{fig:n_exp}b shows the number of gravitational microlensing events expected to be detected by OGLE if the entire dark matter were composed of compact objects of a given mass. The blue and red lines in Figure~\ref{fig:n_exp}b mark the expected number of events that originate from lenses in the Milky Way and LMC haloes, respectively. For PBHs more massive than $0.1\,M_{\odot}$, Milky Way halo lenses contribute to $\approx 70\%$ of the expected events. For lower masses, LMC halo lenses start to dominate, because the LMC events generally have longer timescales than those by Milky Way halo objects and so are easier to detect. The thin solid lines in Figure~\ref{fig:n_exp}b show the contribution from fields observed during both the OGLE-III and OGLE-IV phases (from 2001 to 2020), while dashed lines---fields observed during OGLE-IV only (from 2010 to 2020). The latter contribute less than 10\% of the sensitivity of the experiment, because they are located on the outskirts of the LMC (and therefore contain fewer source stars) and were observed for a shorter period of time.

Given the likelihood function $\mathcal{L}(f, M)$ (Eqs. (\ref{eq:likelihood}) and (\ref{eq:likelihood2})) and our model, we employ the Bayes' theorem to derive the posterior distribution for PBH abundance $P(f | M)$:
\begin{equation}
P(f | M) \propto \mathcal{L}(f, M) P_0(f),
\end{equation}
where $P_0(f)$ is a flat (uniform) prior on $f \in [0, 10]$. We use the Markov chain Monte Carlo (MCMC) sampler of \cite{foreman2013} to sample from the posterior and derive 95\% upper limits on $f$ as a function of the PBH mass $M$. To speed up the calculations, we first evaluate $\ln\mathcal{L}(f,M)$ on a grid of 101 logarithmically spaced masses ranging from $10^{-6}\,M_{\odot}$ to $10^4\,M_{\odot}$ and 101 logarithmically spaced values of $f$ from $10^{-4}$ to $10$, and use the linear interpolation to calculate the likelihood between the grid points.

\subsubsection*{Can all microlensing events detected by OGLE be explained by non-dark-matter objects?}

In this section, we would like to verify our assumption that all gravitational microlensing events detected by OGLE in the direction of the LMC can be explained by known stellar populations. We show that the presented simple models of the LMC and the Milky Way disk are capable of explaining the number and properties (positions, timescales, parallaxes) of virtually all gravitational microlensing events detected in our experiment \cite{gould1994_par,evans2000}. We emphasize that we did not attempt to construct a perfect Milky Way/LMC model nor we did not fit the components of that model to match the data ideally, which would be beyond the scope of this work. Our main goal is to measure the contribution of compact objects to dark matter and study how the choice of an astrophysical model affects our limits on the PBH abundance.

First, we use our fiducial model of the Milky Way disk and the LMC to calculate the theoretical microlensing event rate and the event timescale distribution in each field analyzed by OGLE. Then, we estimate the expected number of events in each field by multiplying the event rate by the duration of the observations, the number of source stars, and the average event detection efficiency (Eq.~(\ref{eq:n_exp})). The results are summarized in Extended Data Table~\ref{tab:self_lensing}, separately for fields observed during both the OGLE-III and OGLE-IV phases (2001--2020) and observed by OGLE-IV only (2010--2020).

We expect to detect 5.7 microlensing events due to self-lensing by stars in the LMC itself and 7.0 or 14.7 events due to stars in the Milky Way disk (assuming disk models by \cite{han_gould2003} or \cite{cautun2020}, respectively). The total mass of the Milky Way disk in the latter model is 2.6 times larger than in the former, which explains why the model predicts twice as many microlensing events. In total, we expect 12.7 or 20.4 events, depending on the Milky Way disk model, which can be compared to the total number of thirteen events in the final statistical sample of \cite{mroz2024a}. The Milky Way disk model of \cite{han_gould2003} is clearly favored, although the model of \cite{cautun2020} is still allowed, because the Poisson probability of observing 13 events given the expected 20.4 or more events is 5.6\%. See Extended Data Figure~\ref{fig:poisson}.

Extended Data Figure~\ref{fig:map} shows the number of events due to known stellar populations in the LMC and Milky Way disk expected to be detected in each OGLE field. Most events are expected to be found in the central regions of the LMC, where the self-lensing event rate is highest and the number of source stars is largest. Approximately 80\% of the events are expected to be located within 3\,deg from the LMC center (defined by \cite{kim1998}) in both the \cite{han_gould2003} and \cite{cautun2020} models. There are two events in the \cite{mroz2024a} sample that are located more than 3\,deg of the LMC center, OGLE-LMC-16 (3.2\,deg) and OGLE-LMC-17 (4.9\,deg), in excellent agreement with the predictions of the \cite{han_gould2003} model (2.5 events). The cumulative distribution of the angular distances of the detected events from the LMC center matches that expected from both models, with $p$-values of 0.29 and 0.16 for the \cite{han_gould2003} and \cite{cautun2020} models, respectively.

The distribution of Einstein timescales of events that are expected to be detected by OGLE (taking into account the detection efficiencies in a given field) is presented in Extended Data Figure~\ref{fig:tE}ab. 80\% of the expected events should have timescales between 26 and 417\,d in the \cite{han_gould2003} model (between 25 and 316\,d in the \cite{cautun2020} model). Events with lenses located in the LMC are expected to have generally longer timescales than those originating from the Milky Way disk lenses, because the relative lens-source proper motions are smaller.

The timescales of the detected events generally match those expected from the model (Extended Data Figure~\ref{fig:tE}ab). However, three events in the sample have timescales shorter than 25\,d (OGLE-LMC-08: $\tE = 13.5^{+6.0}_{-4.0}$\,d, OGLE-LMC-13: $\tE = 7.0^{+2.0}_{-1.1}$\,d, OGLE-LMC-17: $\tE = 13.8^{+1.9}_{-1.1}$\,d), whereas the model predicts that about 10\% of all events (that is, 1.3) should fall in this range. It is possible that this is due to a statistical fluctuation, the Poisson probability of observing three events given the expected 1.3 events is about 10\%. The model may also underpredict the number of low-mass lenses or the number of high-proper motion objects, both of which contribute to the population of short-timescale microlensing events.

Extended Data Figure~\ref{fig:tE}cd shows the distribution of the microlensing parallaxes of events expected to be detected in our experiment, assuming the Milky Way disk models by \cite{han_gould2003} and \cite{cautun2020}. In both cases, the $\piE$ distributions are clearly bimodal: the two peaks correspond to two distinct populations of lenses, those in the Milky Way disk, and those in the LMC \cite{gould1994_par}. The Milky Way disk lenses are located nearby, and so their microlens parallaxes are relatively large; they peak at $\piE \approx 0.5$. The second population of lenses that reside in the LMC is characterized by small parallaxes (typically $\piE \approx 0.01$). The different values of the parallaxes are consistent with our expectations. For the same lens mass, the microlens parallax scales as $\piE \propto \sqrt{\pirel}$, where $\pirel$ is the relative lens--source parallax. For events with Milky Way disk lenses, $\pirel \approx 1/D_{\rm l} \approx 1\,\mathrm{mas}$. For the LMC lenses, $\pirel \approx \Delta D/D_{\rm s}^2 \approx 1/2500\,\mathrm{mas}$, where $\Delta D \approx 1\,\mathrm{kpc}$ is the thickness of the LMC disk, so the LMC self-lensing events should have microlensing parallaxes that are about 50 times smaller than the Milky Way disk events, which is consistent with the simulations presented in Extended Data Figure~\ref{fig:tE}cd.

The statistical sample of \cite{mroz2024a} contains four events with reliably measured microlensing parallaxes. This does not mean that the remaining events have parallaxes close to zero; rather, these events are too short or too faint to robustly measure the microlens parallax. However, whenever the parallax is clearly detected in the light curve and its value is larger than $\piE \approx 0.1$, the lensing object is very likely to be located in the Milky Way disk.

\subsubsection*{Full limits on PBHs}
\label{sec:bounds2}

The Einstein timescales of detected events carry information that can be used to further constrain the abundance of PBHs in the Milky Way and LMC haloes. According to Extended Data Figure~\ref{fig:n_exp}b, if the entire dark matter was composed of black holes of $2200\,M_{\odot}$, we ought to detect thirteen microlensing events. However, the typical Einstein timescales of these events should be on the order of 9\,yr, in stark contrast to the observed values.

To take into account the information included in the timescales of the detected events, we evaluate the full likelihood function defined in Eq.~(\ref{eq:likelihood}). We follow the same procedure as above. We calculate $\ln\mathcal{L}(f,M)$ on a grid of 101 logarithmically spaced masses ranging from $10^{-6}\,M_{\odot}$ to $10^4\,M_{\odot}$ and 101 logarithmically spaced values of $f$ from $10^{-4}$ to $10$.

Although most of the detected events have precisely measured timescales, the fractional error bar on $\tE$ may be larger than 20\% for some events. Thus, we replace the term $\varepsilon(t_{{\rm E},i}) d\Gamma/dt_{\rm E}\left(t_{{\rm E},i},f,M\right)$ in the definition of $\mathcal{L}(f,M)$ by the mean over $N_i$ samples from the posterior distribution of $t_{{\rm E},i}$:
\begin{equation}
 \frac{1}{N_i}\sum_{k=1}^{N_i} \left[ \varepsilon(t_{{\rm E},ik}) \frac{d\Gamma}{dt_{\rm E}}\left(t_{{\rm E},ik},f,M\right) \right],
\end{equation}
where the index $i$ denotes the $i$th event and the index $k$ runs from 1 through $N_i = 2\times 10^5$.

Extended Data Figure~\ref{fig:likelihood} shows the contours of the log-likelihood function, separately for the \cite{han_gould2003} and \cite{cautun2020} models. Regardless of the model chosen, it is clear there is no evidence that PBHs in the mass range $10^{-6}\,M_{\odot} < M_{\rm PBH} < 10^4\,M_{\odot}$ comprise a measurable fraction of the dark matter mass. For the \cite{han_gould2003} model, the likelihood is highest for $(\log M, \log f) = (-1.5, -2.55)$. However, the model with $\log f =-4$ is disfavored by only $\Delta\chi^2 \equiv 2(\ln\mathcal{L}_{\rm max}-\ln\mathcal{L}) \approx 3.7$. Similarly, for the second model, the likelihood is highest for $(\log M, \log f) = (-1.8, -2.85)$, and the model with $\log f =-4$ is disfavored only by $\Delta\chi^2 \approx 1.5$.

We found that the highest-likelihood grid point has $\log f > -4$ mostly due to one event, OGLE-LMC-13, which has the shortest timescale in the sample of only $\tE = 7.0^{+2.0}_{-1.1}$\,d. Indeed, such a short-timescale event seems to be very rare in our simulations (Extended Data Figure~\ref{fig:tE}ab). If the event is removed from the sample, models with $\log f=-4$ are disfavored by less than $\Delta\chi^2 \approx 1.5$.

We calculate the posterior distribution for the PBH abundance $P(f | M)$ using the MCMC approach. Our 95\% upper limits on the PBH abundance in dark matter are presented in Figure~\ref{fig:bounds} by dashed (\cite{han_gould2003} model) and dotted (\cite{cautun2020} model) lines, and they can be compared to those derived above (solid red line in Figure~\ref{fig:bounds}). Overall, the differences are very small for the least massive ($M_{\rm PBH} \lesssim 10^{-4}\,M_{\odot}$) and the most massive ($M_{\rm PBH} \gtrsim 10\,M_{\odot}$) PBHs, because we do not expect that ordinary lenses would produce microlensing events with timescales that could be attributed to those by extremely low- or high-mass PBHs.

In the intermediate mass range ($10^{-4}\,M_{\odot} \lesssim M_{\rm PBH} \lesssim 10\,M_{\odot}$), microlensing events caused by PBHs may be mistaken with those caused by known stellar populations in the LMC or the Milky Way disk. Therefore, the derived limits are slightly weaker. The difference between the 95\% upper limits on PBHs calculated assuming the \cite{han_gould2003} Milky Way disk model and those calculated above is the largest for $\log M = -0.6$ and amounts to $\Delta\log f = 0.59$\,dex (that is, the limits are about a factor of 3.9 weaker). The choice of the Milky Way model may also slightly affect the derived limits. The difference between the 95\% upper limits on PBHs calculated assuming the \cite{han_gould2003} and the \cite{cautun2020} models is greatest for $\log M = -0.1$ and equals to $\Delta\log f = 0.18$\,dex at that mass.

\subsection*{Discussion}

The limits on the PBH abundance that are presented in the previous section (and in Figure~\ref{fig:bounds}) are calculated based on a full statistical sample of thirteen microlensing events detected over 20\,years of the OGLE observations of the LMC \cite{mroz2024a}. As discussed in a companion paper \cite{mroz2024a}, the sample includes all microlensing events with source stars brighter than $I = 22$\,mag. Likewise, the detection efficiencies $\varepsilon(\tE)$ and the number of source stars $N_{\rm s}$ used in the calculations include only sources brighter than $I=22$\,mag. However, the event detection efficiency rapidly drops with the brightness of source stars; the fainter the source, the lower the chances of detecting the microlensing event. Faint source stars are additionally more difficult to count because of the increased amount of blending from unrelated stars \cite{mroz2024a}. As a result, although the final sample of events is larger, there is a risk of introducing unknown systematic errors in the analysis.

Events with brighter source stars should be less prone to such systematic effects. Therefore, we check how our results depend on the adopted limiting magnitude. We re-calculate the detection efficiency for all fields and the number of source stars for limiting magnitudes of $I=21$ and $I=21.5$\,mag. Then, we follow the methods described above to infer the 95\% upper limits on the PBH abundance. The results of the calculations are presented in Extended Data Figure~\ref{fig:comparison}b. Red and blue lines in Extended Data Figure~\ref{fig:comparison}b mark the limits for the limiting magnitudes of 21 and 21.5\,mag, respectively, whereas black lines indicate the original limits calculated for the limiting magnitude of 22\,mag.

As expected, the differences between the inferred limits are relatively small. The limits for the limiting magnitudes of 21.5 and 22\,mag differ by at most $\Delta\log f=-0.10$\,dex, for the limiting magnitudes of 21 and 22\,mag---by at most $\Delta\log f=-0.20$\,dex. This demonstrates that possible systematic effects related to the inclusion of faint source stars are very small and do not influence our conclusions.

The recent studies of the Galactic rotation curve by \cite{jiao2023} and \cite{ou2024} seem to indicate that the mass of the dark matter Milky Way halo may be smaller than in the adopted model. In the first-order approximation, our limits on PBHs are inversely proportional to the microlensing optical depth (the larger the optical depth, the stronger the limits). The microlensing optical depth toward the center of the LMC in our fiducial Milky Way dark matter halo model \cite{cautun2020} is $4.4 \times 10^{-7}$. If we use the ``B2'' halo model of \cite{jiao2023} (with the Einasto profile of index 0.43), then the optical depth toward the center of the LMC is approximately $4.2 \times 10^{-7}$, and our limits on PBHs as dark matter are very similar to those calculated with the fiducial model (Extended Data Figure~\ref{fig:comparison}a). For the best-fit model from \cite{ou2024} (Einasto profile of index 1.1), the optical depth is $4.1\times 10^{-7}$.

Because the distribution of dark matter predicted by halo models of \cite{jiao2023} and \cite{ou2024} is more compact than in our fiducial model, the mean distance to the lens (in the direction of the LMC center) is shorter: $\langle D_l \rangle = 7.6$\,kpc \cite{jiao2023}, 8.5\,kpc \cite{ou2024}, 12.5\,kpc \cite{cautun2020}. Therefore, both angular Einstein radii and relative lens--source proper motions are larger. These two effects nearly cancel each other out, and the resulting mean Einstein timescales and event rates are very similar in all three models considered \cite{cautun2020,jiao2023,ou2024}.

\subsection*{LMC proper motion and rotation}
\label{app:lmc_kin}

We use the \textit{Gaia}~EDR3 data \cite{gaia2016,gaia_edr3} to measure the mean LMC proper motion and to devise a simple model of the LMC rotation. The proper motion and rotation of the LMC (and other nearby dwarf galaxies) were extensively studied by \cite{helmi2018} based on \textit{Gaia}~DR2 data and here we closely follow their approach. We select stars brighter than $G=19$ that are located within 8\,deg of the dynamical center of the H\textsc{i} LMC disk ($\alpha_0=78.77^{\circ}$, $\delta_0=-69.01^{\circ}$) \cite{van_der_marel2014,helmi2018}, have a reliable astrometric solution (with $\textrm{RUWE} \leq 1.4$), and have parallaxes consistent with that of the LMC ($\varpi \leq 1$\,mas and $\varpi/\sigma_{\varpi} \leq 10$), in total about 4.8~million sources. We then calculate the positions $x$, $y$ and proper motions $\mu_x$, $\mu_y$ of all stars in an orthographic projection centered at $(\alpha_0, \delta_0)$ using equations:
\begin{align}
\begin{split}
x =& \cos\delta\sin(\alpha-\alpha_0), \\
y =& \sin\delta\cos\delta_0-\cos\delta\sin\delta_0\cos(\alpha-\alpha_0),\\
\mu_x =& \mu_{\alpha}\cos(\alpha-\alpha_0)-\mu_{\delta}\sin\delta\sin(\alpha-\alpha_0),\\
\mu_y =& \mu_{\alpha}\sin\delta_0\sin(\alpha-\alpha_0)\\
 &+ \mu_{\delta}(\cos\delta\cos\delta_0+\sin\delta\sin\delta_0\cos(\alpha-\alpha_0)).
\end{split}
\end{align}
The median proper motions of the selected stars are presented in Extended Data Figure~\ref{fig:lmc_pm}.

The median proper motion of the LMC (which is determined using stars located within $5^{\circ}$ of the dynamical center) is $\mu_x = 1.851 \pm 0.340$\,mas\,yr$^{-1}$ and $\mu_y = 0.277 \pm 0.382$\,mas\,yr$^{-1}$, in good agreement with the results of \cite{helmi2018}. Here the uncertainties denote the dispersion of proper motions in a given direction. This corresponds to $\mu_l = -0.662 \pm 0.381$\,mas\,yr$^{-1}$ and $\mu_b = 1.751 \pm 0.342$\,mas\,yr$^{-1}$ in the Galactic coordinates.

The proper motion pattern seen in the middle and bottom panels of Extended Data Figure~\ref{fig:lmc_pm} reflects the rotation of the LMC and, in the first order, can be approximated by the central value and two gradients.  Additional striping pattern of lower amplitude is indicative of small-scale systematics in \textit{Gaia}~EDR3 \citep{lindegren2021}. To minimize their influence, we calculate the median proper motion (and its dispersion) in $100 \times 100$ bins in the range $|x| \leq 0.15$\,rad and $|y| \leq 0.15$\,rad and fit the following linear model to the binned data:
\begin{align}
\begin{split}
\mu_x = \mu_{x,0} + \frac{\partial\mu_x}{\partial x}x + \frac{\partial\mu_x}{\partial y}y, \\
\mu_y = \mu_{y,0} + \frac{\partial\mu_y}{\partial x}x + \frac{\partial\mu_y}{\partial y}y.
\end{split}
\end{align}
The six fit coefficients are determined using the least squares approach using bins with at least 10 stars. The model is fit in 7 annuli with a width of 1\,deg. The best-fit parameters are presented in Extended Data Table~\ref{tab:lmc_pm}.

\section*{Data availability}

The data used to perform the analysis (event detection efficiencies, source star counts, posterior distributions of event parameters) are publicly available at \url{https://www.astrouw.edu.pl/ogle/ogle4/LMC_OPTICAL_DEPTH/} and Zenodo (\url{https://zenodo.org/doi/10.5281/zenodo.10879576}). 

\section*{Code availability}

The custom codes for the simulation of microlensing events toward the LMC and the calculation of limits on PBHs as dark matter are available upon request from the corresponding author.

\makeatletter
\apptocmd{\thebibliography}{\global\c@NAT@ctr 30\relax}{}{}
\makeatother

\section*{Acknowledgements}

We thank all the OGLE observers for their contribution to the collection of the photometric data over the decades. We thank T. Bulik for insightful comments on the manuscript. This research was funded in part by National Science Centre, Poland, grant OPUS 2021/41/B/ST9/00252 awarded to P.M. 

\section*{Author contributions}

P.M. led the analysis and interpretation of the data and wrote the manuscript. A.U. is the PI of the OGLE project, and was responsible for the data reduction. All authors collected the OGLE photometric observations, and reviewed, discussed, and commented on the presented results and on the manuscript.

\section*{Competing interests}

The authors declare no competing interests.

\section*{Additional information}

\textbf{Correspondence and requests for materials} should be addressed to Przemek Mr\'oz (pmroz@astrouw.edu.pl).

\noindent \textbf{Reprints and permissions information} is available at http://www.nature.com/reprints.

\renewcommand{\figurename}{Extended Data Figure}
\renewcommand{\tablename}{Extended Data Table}
\setcounter{figure}{0}    
\setcounter{table}{0}    

\newpage
\clearpage

\begin{figure}
\centering
\includegraphics[width=.9\textwidth]{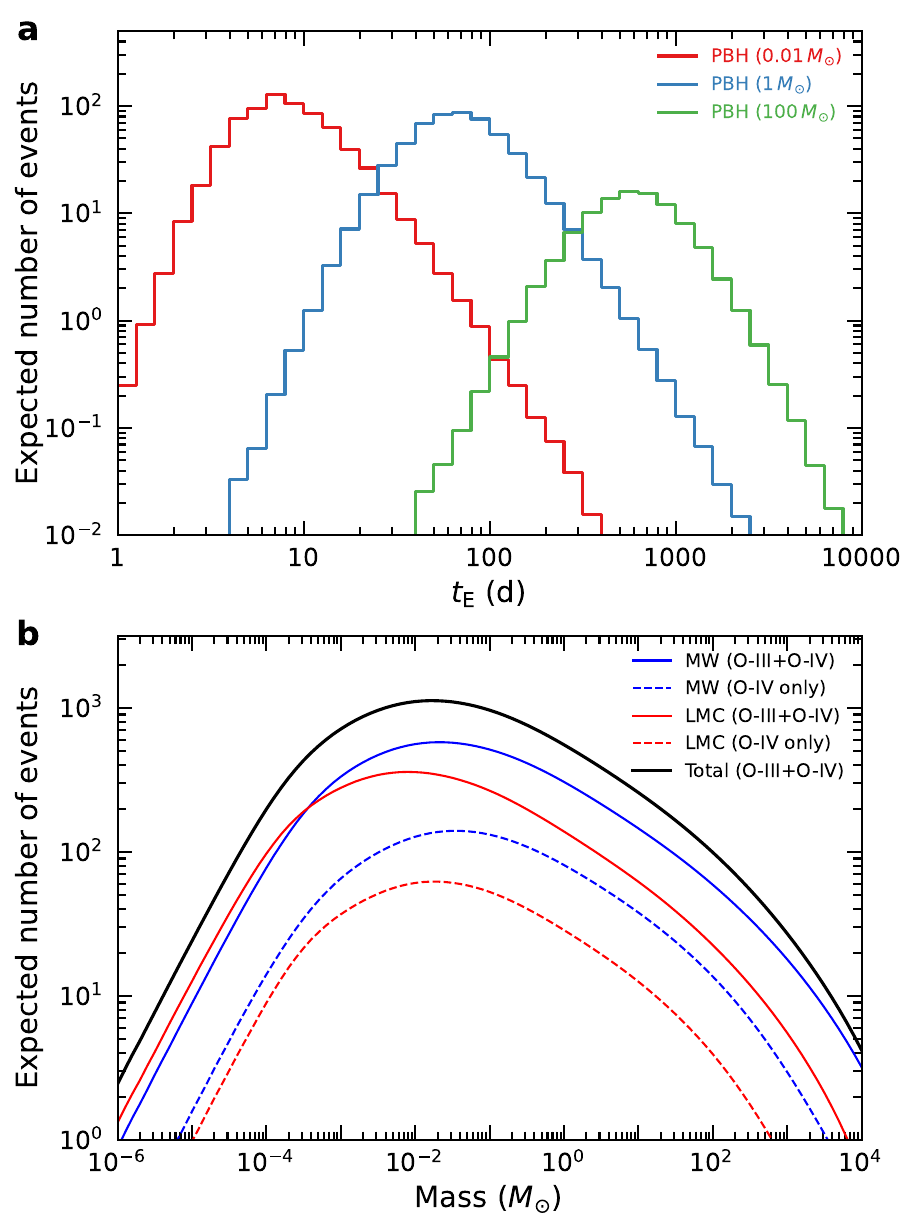}
\caption{\textbf{Expected number of microlensing events from a PBH dark matter halo.} \textbf{a,} Number of microlensing events as a function of their Einstein timescale that should have been discovered by OGLE assuming that the entire dark matter were composed of PBHs of 0.01 (red), 1 (blue), and $100\,M_{\odot}$ (green). \textbf{b,} Number of gravitational microlensing events expected to be detected by OGLE if entire dark matter were composed of compact objects of a given mass $N_{\rm exp}({f=1},M)$. Thin solid lines correspond to fields observed during OGLE-III and OGLE-IV phases (from 2001 to 2020), dashed lines -- fields observed during OGLE-IV only (from 2010 to 2020). Blue lines mark the contribution from the Milky Way dark matter halo, red lines -- the LMC dark matter halo.}
\label{fig:n_exp}
\end{figure}

\begin{figure}
\centering
\includegraphics[width=.75\textwidth]{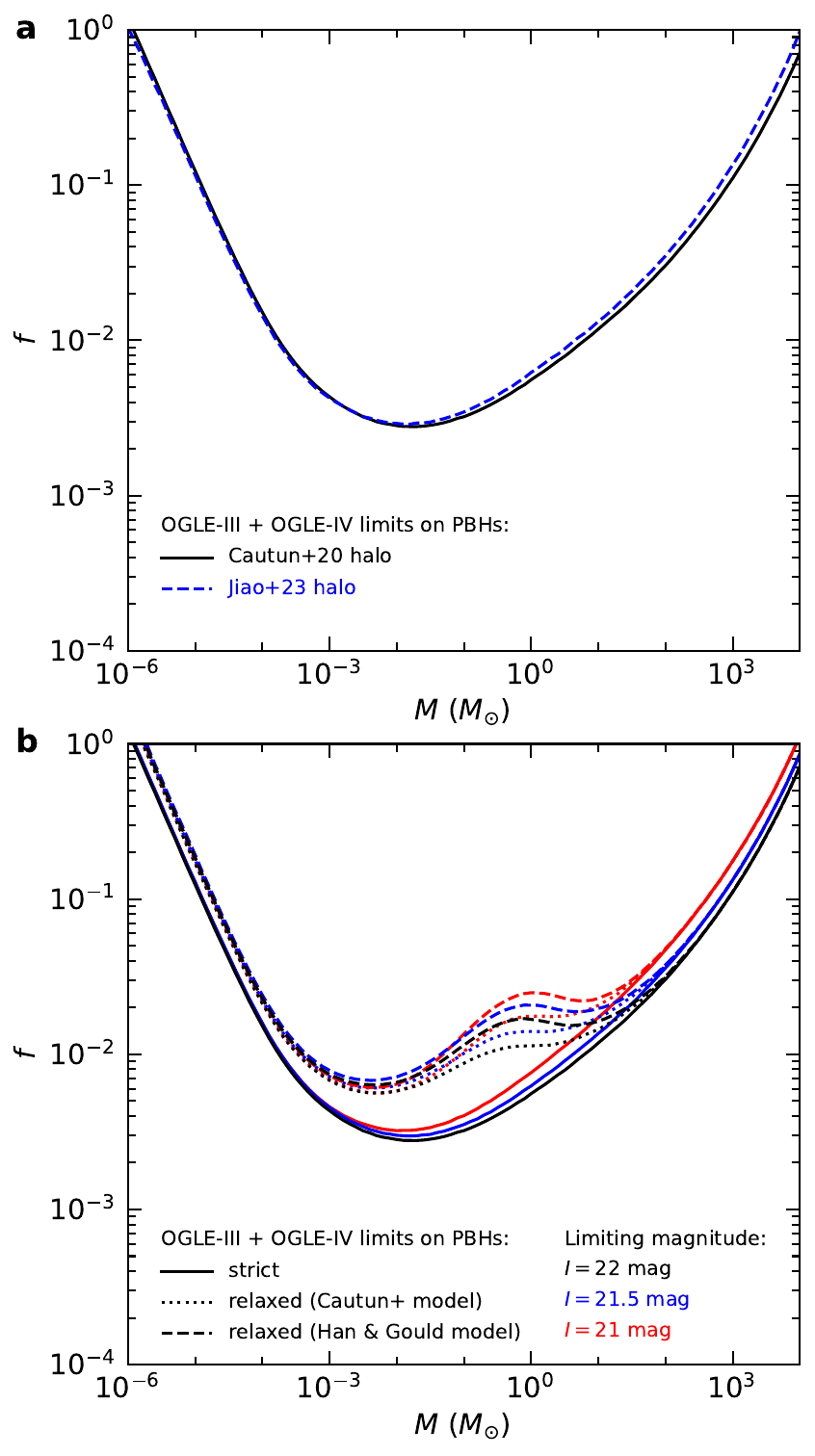}
\caption{\textbf{95\% upper limits on PBHs as constituents of dark matter.} \textbf{a,} Dependence of the limits on the Milky Way halo model. The black solid line marks limits for the Cautun et al. (2020) \cite{cautun2020} model. The blue dashed line---Jiao et al. (2023) \cite{jiao2023} model. \textbf{b}, Limits as a function of the limiting magnitude.}
\label{fig:comparison}
\end{figure}

\begin{figure}
\centering
\includegraphics[width=\textwidth]{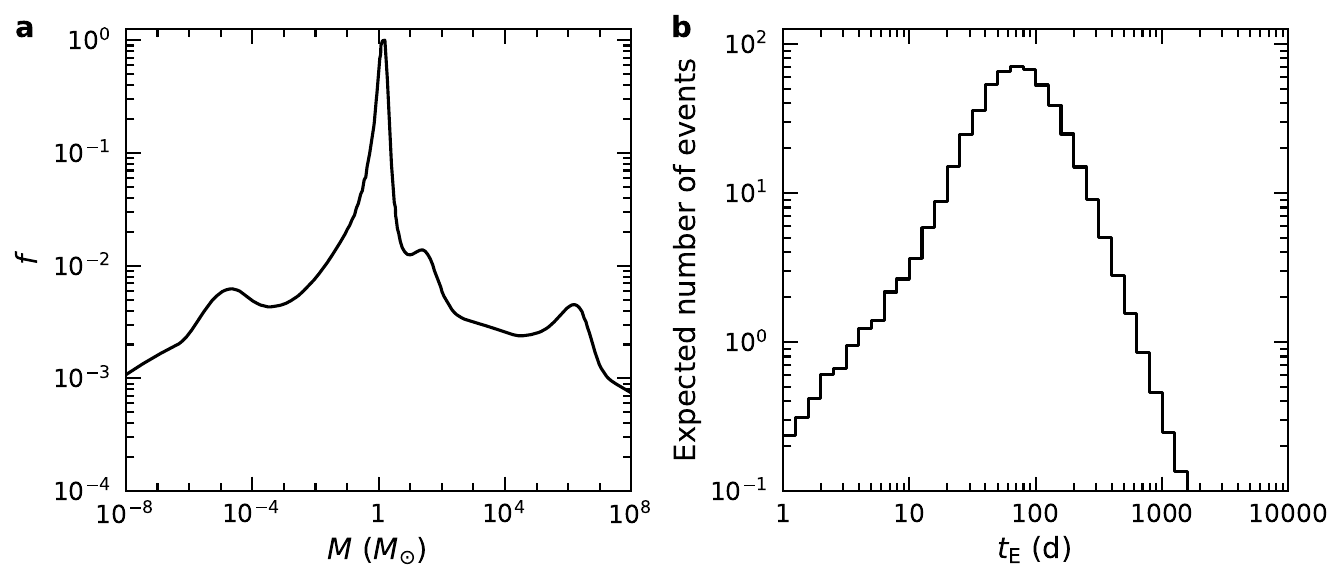}
\caption{ \textbf{Predictions for a multi-peak PBH mass function.} \textbf{a,} Multi-peak mass function of PBHs from \cite{carr2021b} (assuming spectral index $\tilde{n}_{\rm s}=0.960$). \textbf{b,} Expected distribution of event timescales assuming that the entire dark matter is composed of PBHs with the mass function from \cite{carr2021b}. In total, we should have detected 513 microlensing events.}
\label{fig:carr_model}
\end{figure}

\begin{figure}
\centering
\includegraphics[width=\textwidth]{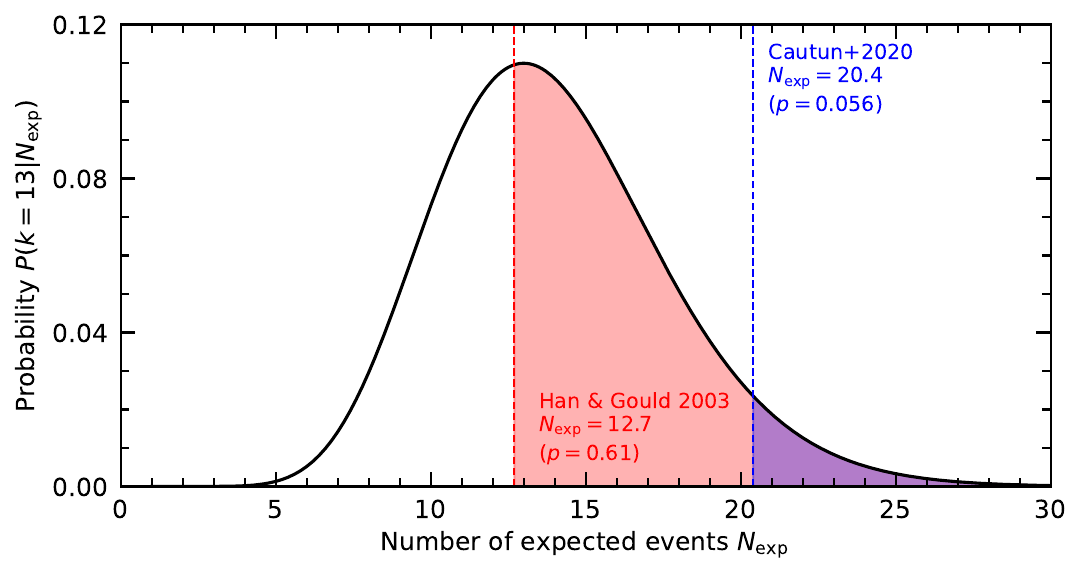}
\caption{\textbf{Poisson probability of observing 13 events as a function of the expected number of events in the model.} The blue and red dashed lines mark the models by \cite{cautun2020} and \cite{han_gould2003}, respectively.}
\label{fig:poisson}
\end{figure}

\begin{figure}
\includegraphics[width=\textwidth]{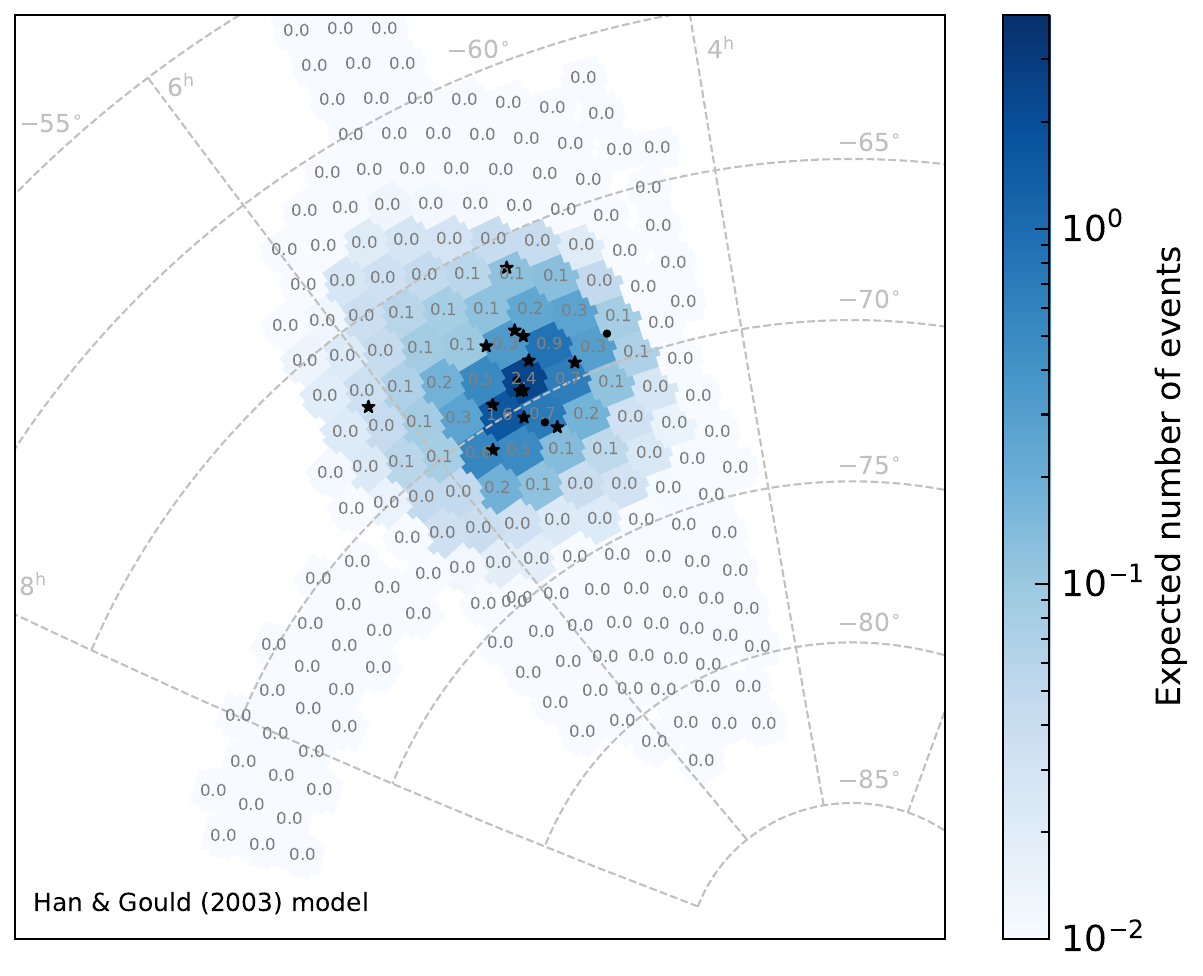}
\caption{ \textbf{Expected number of microlensing events from known stellar populations in the LMC and Milky Way disk in OGLE-IV fields.} The Milky Way disk model used in calculations is from Han \& Gould (2003) \cite{han_gould2003}. Black asterisks mark events that are a part of the statistical sample of \cite{mroz2024a}, black dots mark other events.}
\label{fig:map}
\end{figure}

\begin{figure*}
\includegraphics[width=\textwidth]{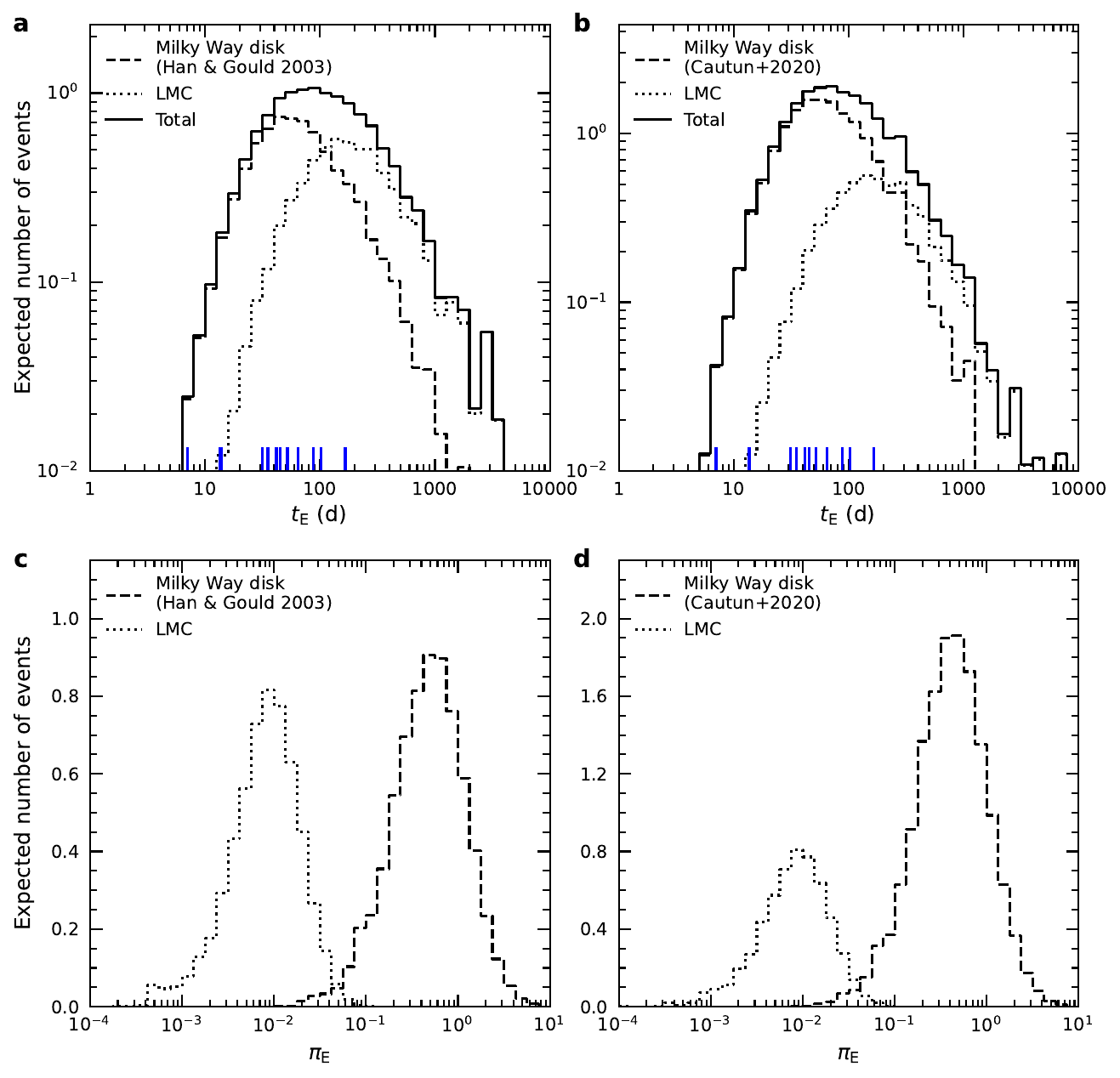}
\caption{\textbf{Expected distributions of timescales and parallaxes of microlensing events from known stellar populations in the LMC and Milky Way disk.} \textbf{a, b,} Expected distributions of Einstein timescales for the Han \& Gould (2003) \cite{han_gould2003} and Cautun et al. (2020) \cite{cautun2020} Milky Way disk models. Vertical blue lines mark the detected events. \textbf{c, d,} Expected distributions of microlensing parallaxes.}
\label{fig:tE}
\end{figure*}

\begin{figure*}[t]
\centering
\includegraphics[width=0.8\textwidth]{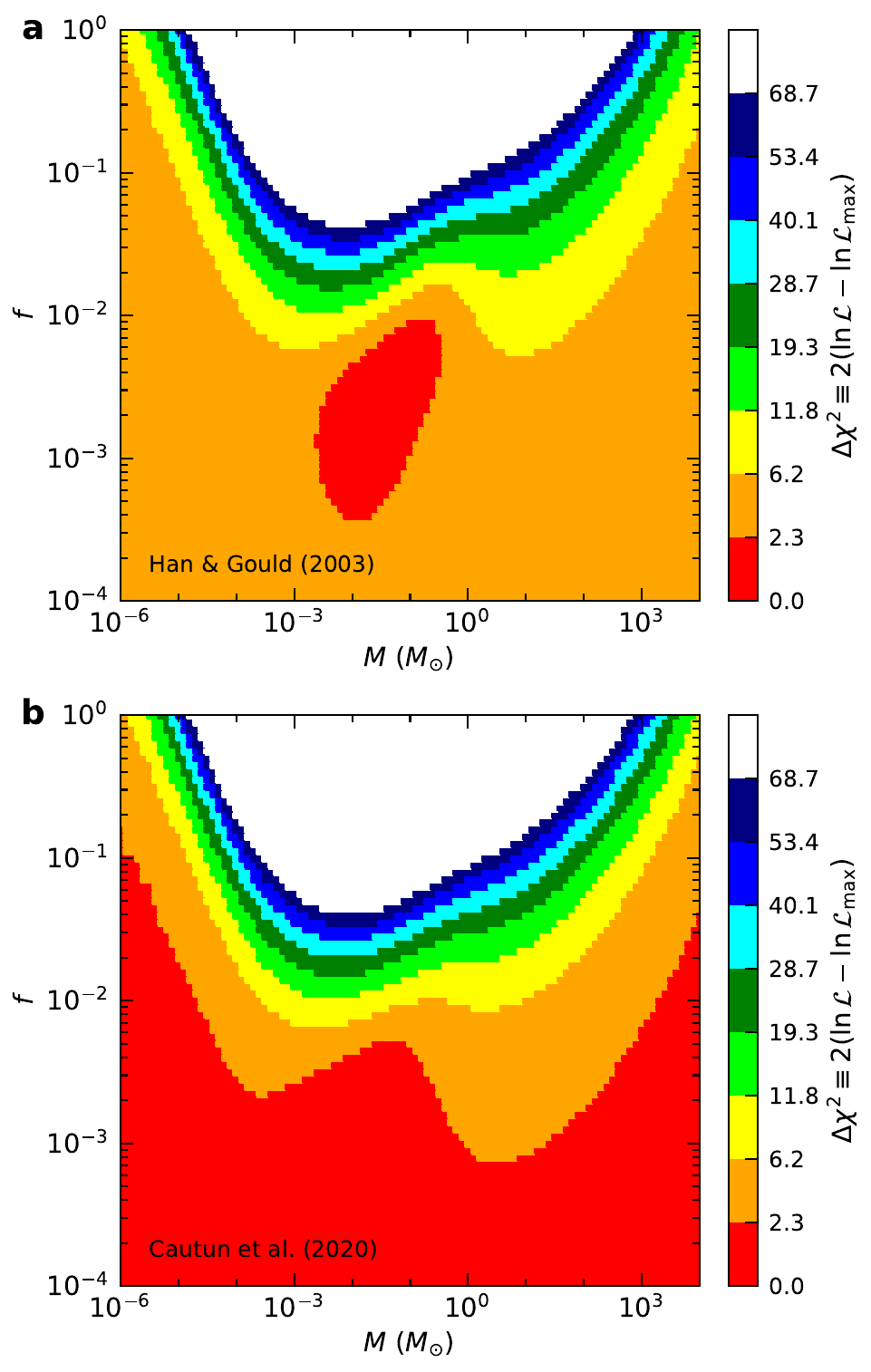}
\caption{\textbf{Contours of the likelihood function (Eq.~\ref{eq:likelihood}).} \textbf{a,} Contours for the Han \& Gould (2003) \cite{han_gould2003} Milky Way disk model. \textbf{b,} Contours for the Cautun et al. (2020) \cite{cautun2020} Milky Way disk model. The color codes the difference $\Delta\chi^2 \equiv 2(\ln\mathcal{L}_{\rm max}-\ln\mathcal{L})$. }
\label{fig:likelihood}
\end{figure*}

\begin{figure*}
\centering
\includegraphics[width=0.8\textwidth]{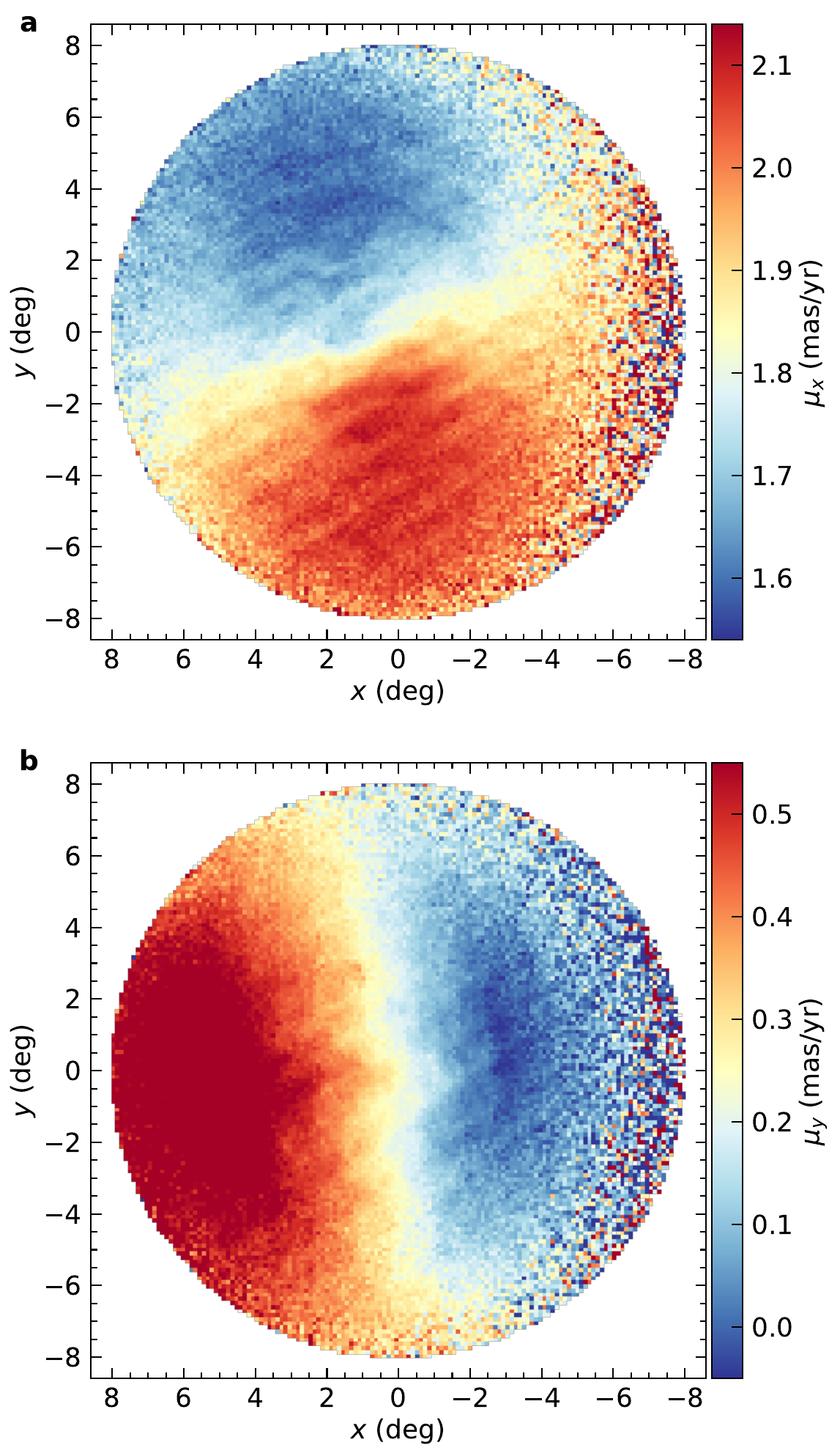}
\caption{\textbf{\textit{Gaia}~EDR3 proper motions of stars in the LMC.}}
\label{fig:lmc_pm}
\end{figure*}

\newpage
\clearpage

\begin{table*}
\begin{tabular}{lccc}
\hline \hline
& OGLE-III \& OGLE-IV & OGLE-IV Only & Total \\
\hline
Milky Way disk \cite{han_gould2003} &  5.5 & 1.5 &  7.0 \\
Milky Way disk \cite{cautun2020}    & 11.5 & 3.2 & 14.7 \\
LMC                                  &  5.3 & 0.4 &  5.7 \\
\hline
Total (disk model \cite{han_gould2003})          & 10.8 & 1.9 & 12.7 \\
Total (disk model \cite{cautun2020})             & 16.8 & 3.6 & 20.4 \\
\hline
\end{tabular}
\caption{\textbf{Expected number of microlensing events from known stellar populations in the Milky Way disk and the LMC.} We separately list contributions from fields observed by both OGLE-III and OGLE-IV (second column) and OGLE-IV only (third column).}
\label{tab:self_lensing}
\end{table*}

\begin{table*}
\scriptsize
\begin{tabular}{ccccccccc}
\hline \hline
$\rho$ (deg)& $\partial\mu_x/\partial x$ & $\partial\mu_x/\partial y$ & $\mu_{x,0}$ & $\partial\mu_y/\partial x$ & $\partial\mu_y/\partial y$ & $\mu_{y,0}$ & rms($\mu_x$) & rms($\mu_y$)\\
\hline
$0 < \rho < 1$ & $-3.6832$ & $-6.7018$ & 1.8535 & 5.6324 & $-1.5691$ & 0.2348 & 0.029 & 0.023 \\
$1 < \rho < 2$ & $-2.3530$ & $-5.6825$ & 1.8565 & 5.7179 & $-0.7549$ & 0.2135 & 0.041 & 0.027 \\
$2 < \rho < 3$ & $-1.2292$ & $-4.5601$ & 1.8471 & 5.3498 & $-0.3510$ & 0.2178 & 0.037 & 0.028 \\
$3 < \rho < 4$ & $-1.0871$ & $-3.5981$ & 1.8355 & 4.5304 & $-0.5892$ & 0.2357 & 0.035 & 0.036 \\
$4 < \rho < 5$ & $-0.9070$ & $-2.7004$ & 1.8348 & 3.7424 & $-0.5745$ & 0.2659 & 0.042 & 0.057 \\
$5 < \rho < 6$ & $-0.7520$ & $-2.1700$ & 1.8430 & 3.2343 & $-0.4711$ & 0.2770 & 0.049 & 0.058 \\
$6 < \rho < 7$ & $-0.7497$ & $-1.7328$ & 1.8510 & 2.8296 & $-0.4742$ & 0.2655 & 0.048 & 0.050 \\
\hline
\end{tabular}
\caption{\textbf{Parameters of the LMC proper motion model.} Units of $\partial\mu_x/\partial x$, $\partial\mu_x/\partial y$, $\partial\mu_y/\partial x$, and $\partial\mu_y/\partial y$ are mas\,yr$^{-1}$\,rad$^{-1}$, whereas $\mu_{x,0}$, $\mu_{y,0}$, rms($\mu_x$), and rms($\mu_y$) are given in mas\,yr$^{-1}$.}
\label{tab:lmc_pm}
\end{table*}



\end{document}